\documentclass[preprint,aps,showpacs,preprintnumbers,amsmath,amssymb]{revtex4}  

\usepackage{graphicx}
\usepackage{dcolumn}
\usepackage{bm}

\begin{document}


\title{Frequency-Domain Coherent Control of Femtosecond Two-Photon Absorption: Intermediate-Field vs. Weak-Field Regime}

\author{Lev Chuntonov, Leonid Rybak, Andrey Gandman, and Zohar Amitay}
\email{amitayz@tx.technion.ac.il} %
\affiliation{Schulich Faculty of Chemistry, Technion - Israel
Institute of Technology, Haifa 32000, Israel}

\begin{abstract}
Coherent control of femtosecond two-photon absorption in the
intermediate-field regime is analyzed in detail in the powerful
frequency domain using an extended 4$^{th}$-order perturbative
description. The corresponding absorption is coherently induced by
the weak-field non-resonant two-photon transitions as well as by
four-photon transitions involving three absorbed photons and one
emitted photons. The interferences between these two groups of
transitions lead to a difference between the intermediate-field and
weak-field absorption dynamics. The corresponding interference
nature (constructive or destructive) strongly depends on the
detuning direction of the pulse spectrum from half the two-photon
transition frequency. The model system of the study is atomic
sodium, for which both experimental and theoretical results are
obtained. The detailed understanding obtained here serves as a basis
for coherent control with rationally-shaped femtosecond pulses in a
regime of sizable absorption yields.
\end{abstract}


\maketitle

\section{\label{sec:introduction}Introduction}

Femtosecond pulses offer unique ways to coherently control photo-induced quantum dynamics of matter   
\cite{tannor_kosloff_rice_coh_cont,shapiro_brumer_coh_cont_book,warren_rabitz_coh_cont,
rabitz_vivie_motzkus_kompa_coh_cont,dantus_exp_review1_2}.
The corresponding key characteristic is their coherence over a broad spectrum.
Multiphoton absorption processes in atoms and molecules, which are
of fundamental scientific importance as well as applicative
importance to the fields of nonlinear spectroscopy and microscopy,
are among the processes that have been controlled  most effectively
by shaped femtosecond pulses
\cite{dantus_exp_review1_2,silberberg_2ph_nonres1_2,dantus_2ph_nonres_molec1_2,
baumert_2ph_nonres,silberberg_2ph_1plus1,girard_2ph_1plus1,becker_2ph_1plus1_theo,silberberg_antiStokes_Raman_spect,
gersh_murnane_kapteyn_Raman_spect,leone_res_nonres_raman_control,
amitay_3ph_2plus1,amitay_2ph_inter_field1,
silberberg-2ph-strong-field,weinacht-2ph-strong-theo-exp,wollenhaupt-baumert1_2,hosseini_theo_multiphoton_strong}.
The control principle is the coherent manipulation of interferences among
the manifold of initial-to-final state-to-state multiphoton pathways that are induced by the pulse.
Constructive interferences lead to absorption enhancement (i.e., increased transition probability),
while destructive interferences lead to absorption attenuation (i.e., decreased transition probability).
The interference manipulation is implemented by shaping the femtosecond pulse \cite{pulse_shaping}, i.e.,
manipulating the spectral phase, amplitude, and/or polarization of its different frequency components.
Hence, in order to fully utilize the coherent control potential of a given excitation scheme,
the ideal line of action is shaping the pulse based on an initial identification of the
different multiphoton pathways and their interference mechanism.
When such identification is not possible, a practical partial solution is to use automatic experimental optimization
of the pulse shape using learning algorithms that generally considering the system as a black box
\cite{rabitz_feedback_learning_idea}.
The lesson, which can be learned from the many successful coherent control studies of multiphoton absorption
conducted in the past
\cite{dantus_exp_review1_2,silberberg_2ph_nonres1_2,dantus_2ph_nonres_molec1_2,
baumert_2ph_nonres,silberberg_2ph_1plus1,girard_2ph_1plus1,becker_2ph_1plus1_theo,
silberberg_antiStokes_Raman_spect,gersh_murnane_kapteyn_Raman_spect,leone_res_nonres_raman_control,
amitay_3ph_2plus1,amitay_2ph_inter_field1},
is that this ideal line of action is feasible and very powerful once the photo-excitation
picture is available in the frequency domain.
This is possible only within the framework of perturbation theory,
where a valid perturbative description in the time domain is (Fourier) transformed
to the frequency domain.
However, until recently the frequency domain has been exploited only in the weak-field regime
\cite{dantus_exp_review1_2,silberberg_2ph_nonres1_2,dantus_2ph_nonres_molec1_2,
baumert_2ph_nonres,silberberg_2ph_1plus1,girard_2ph_1plus1,becker_2ph_1plus1_theo,
silberberg_antiStokes_Raman_spect,gersh_murnane_kapteyn_Raman_spect,leone_res_nonres_raman_control,amitay_3ph_2plus1},
where the N-photon absorption is described by perturbation theory of the lowest
non-vanishing order, i.e., the N$^{th}$ order.
Physically it means that the N-photon absorption is coherently induced 
by all the possible initial-to-final state-to-state pathways of N absorbed photons.
For two-photon absorption the lowest order is the 2$^{nd}$ one, involving all the pathways of two absorbed photons
\cite{dantus_exp_review1_2,silberberg_2ph_nonres1_2,dantus_2ph_nonres_molec1_2,
baumert_2ph_nonres,silberberg_2ph_1plus1,girard_2ph_1plus1,becker_2ph_1plus1_theo}.
The downside of being limited to the weak-field regime is the low absorption yields associated with it.
For two-photon absorption they are typically below 0.1\% population transfer.

Recently \cite{amitay_2ph_inter_field1}, we have extended the powerful frequency-domain picture
of femtosecond two-photon absorption to a regime of considerable absorption yields,
exceeding the weak-field yields by more than two orders of magnitude.
It corresponds to intermediate field strengths, where
the interfering pathways are the weak-field (non-resonant) pathways of two absorbed photons
as well as additional four-photon pathways of three absorbed photons and one emitted photon.
The picture is based on 4$^{th}$-order perturbation theory, which includes both the 2$^{nd}$ and
4$^{th}$ orders associated, respectively, with the two- and four-photon pathways. The relative contribution
of the 4$^{th}$-order absorption amplitude increases as the field strength (pulse intensity) increases.
This intermediate-field regime is distinguished from the strong-field
regime where no perturbative description is valid. The strong-field regime is actually the one that all the
other past multiphoton control studies, which have deviated from the weak-field regime, have focused on
\cite{silberberg-2ph-strong-field,weinacht-2ph-strong-theo-exp,wollenhaupt-baumert1_2,hosseini_theo_multiphoton_strong}.
Our previous work \cite{amitay_2ph_inter_field1} has focused 
on the family of spectral phase patterns that are
anti-symmetric around half the two-photon transition frequency ($\omega_{fg}/2$).
We have found this family to enhance the intermediate-field two-photon absorption relative to the unshaped transform-limited pulse,
when the central spectral frequency is properly detuned, to the red or to the blue
(depending on the system), from $\omega_{fg}/2$.

In the present work we systematically study in detail the
intermediate-field coherent control and the corresponding
interference mechanisms, including their dependence on the pulse
spectrum and its detuning from $\omega_{fg}/2$.
The absorption dynamics in the intermediate-field regime is compared
with the one in the weak-field regime.
The model system is the sodium (Na) atom. As a test case for
femtosecond phase control the study uses the family of shaped pulses
having a $\pi$ spectral phase step, which in the weak-field regime
allows high degree of control over the full accessible range of the
non-resonant two-photon absorption.
Section~\ref{sec:theoretical} presents and elaborates on the extended frequency-domain
4$^{th}$-order perturbative theoretical description. 
The Na intermediate-field control results are presented in Sec.~\ref{sec:results}. They include
experimental results, exact non-perturbative results calculated by the numerical propagation of the
time-dependent Schr\"{o}dinger equation,
and perturbative results calculated numerically using the frequency-domain 4$^{th}$-order formulation.
The formers are used to validate the latter.
Then, in Sec.~\ref{sec:discussion}, the perturbative results are
analyzed and discussed based on their corresponding frequency-domain
description, which allows the identification of the interference
mechanisms leading to the
different intermediate-field features. 

\section{\label{sec:theoretical}Intermediate-Field Theoretical Description}

\begin{figure} 
\includegraphics[width=17cm]{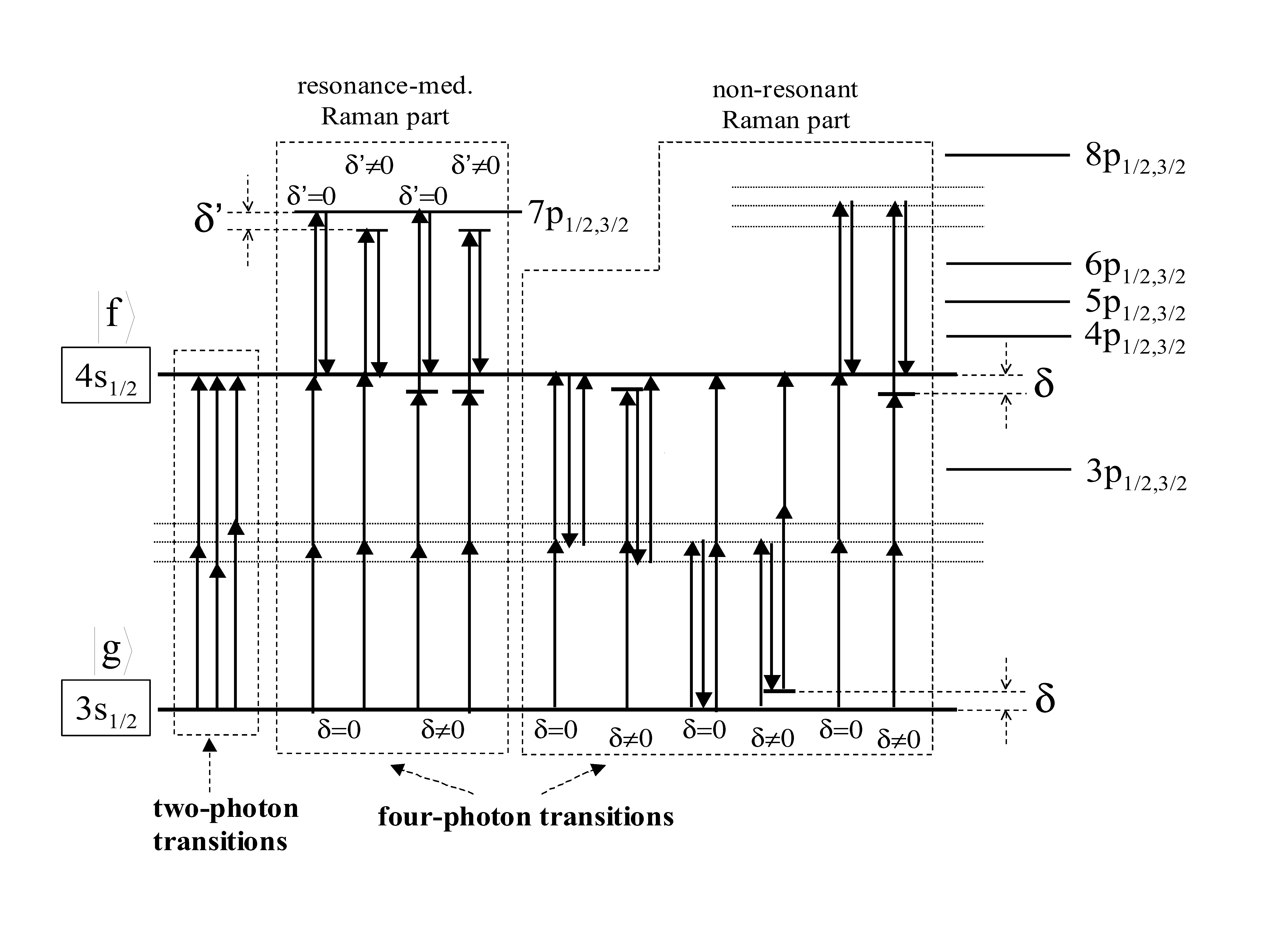}     
\vspace*{-0.7cm}
\caption{Excitation scheme of femtosecond two-photon absorption in the
intermediate-field regime. The indicated levels correspond to the Na atom (not to scale).
Shown are pathway examples of non-resonant two-photon transitions and four-photon transitions
from $\left| f \right\rangle \equiv 4s$ to $\left| g \right\rangle \equiv 3s$. 
The four-photon transitions involve three absorbed photons and one emitted photon in any possible order,
and thus can be decomposed into two parts: a non-resonant two-photon transition and a Raman transition.
The border line between these two parts can be either on-resonance or near-resonance with $3s$ or $4s$
(with detuning $\delta$).
The Raman transition itself can be non-resonant due to the $np$ states (except for $7p$) or on/near-resonance with $7p$ (with detuning $\delta'$).
} \label{fig_1}
\end{figure}

The atomic femtosecond two-photon absorption process we consider is from an
initial ground state $\left|g\right>$ to a final excited state
$\left|f\right>$, which are coupled via a manifold 
states $\left|n\right>$ having the proper symmetry. The spectrum of
the pulse is such that all the
$\left|g\right>$-$\left|n\right>$ and
$\left|f\right>$-$\left|n\right>$ couplings are non-resonant, i.e.,
the spectral amplitude at all the corresponding transition
frequencies is zero: $\left|E(\omega_{gn})\right| =
\left|E(\omega_{fn})\right| = 0$, except for the
$\left|f\right>$-$\left|n_{r}\right>$ resonant coupling for which in
general $\left|E(\omega_{fn_{r}})\right| \ne 0$. The corresponding
excitation scheme is shown schematically in Fig.~\ref{fig_1}.

Within the present intermediate-field regime, the time-dependent (complex) amplitude
$a_{f}(t)$ of state $\left|f\right>$ at time $t$, following irradiation with a
(shaped) temporal electric field $\varepsilon(t)$, can be validly
described by 4$^{th}$-order time-dependent perturbation theory.
So, in general, it includes non-vanishing contributions
from both the 2$^{nd}$ and 4$^{th}$ perturbative orders:
\begin{equation}\label{eq1:time-amp}
a_{f}(t) = a_{f}^{(2)}(t) + a_{f}^{(4)}(t) \; ,
\end{equation}
with
\begin{eqnarray}
a_{f}^{(2)}(t) &=& -\frac{1}{\hbar^{2}}\sum_{m}\mu_{fm}\mu_{mg}
\int_{-\infty}^{t}\int_{-\infty}^{t_{1}}
\varepsilon(t_{1})\varepsilon(t_{2})
\exp\left[i(\omega_{fm}t_{1}+\omega_{mg}t_{2})\right]dt_{1}dt_{2}
\label{eq2:time-amp} \; , \\
a_{f}^{(4)}(t) &=&
-\frac{1}{\hbar^{4}}\sum_{k,l,m}\mu_{fk}\mu_{kl}\mu_{lm}\mu_{mg}
\int_{-\infty}^{t}\int_{-\infty}^{t_{1}}\int_{-\infty}^{t_{2}}\int_{-\infty}^{t_{3}}
\varepsilon(t_{1})\varepsilon(t_{2})\varepsilon(t_{3})\varepsilon(t_{4}) \nonumber \\
& & \times
\exp\left[-i(\omega_{fk}t_{1}+\omega_{kl}t_{2}+\omega_{lm}t_{3}+\omega_{mg}t_{4})\right]
dt_{1}dt_{2}dt_{3}dt_{4} \;, \label{eq3:time-amp}
\end{eqnarray}
%
where $\mu_{ij}=\left<i\right|\hat{\mu}\left|j\right>$ is the
dipole matrix element between a pair of states and
$\omega_{ij}=(E_{i}-E_{j})/\hbar$ is the corresponding transition frequency.
The 2$^{nd}$-order term $a_{f}^{(2)}(t)$ by itself corresponds to the weak-field regime.
The intermediate-field final amplitude $A_{f} \equiv
a_{f}(t\rightarrow\infty)$ of state $\left|f\right>$ after the pulse
is over (i.e., $t\rightarrow\infty$) can be expressed as
\begin{equation}
A_{f} = A_{f}^{(2)} + A_{f}^{(4)} \; ,
\label{eq4:time-amp}
\end{equation}
with $A_{f}^{(2)} \equiv a_{f}^{(2)}(t\rightarrow\infty)$ and $A_{f}^{(4)} \equiv a_{f}^{(4)}(t\rightarrow\infty)$.
The final $\left|f\right>$ population $P_{f} = \left|A_{f}\right|^{2} = [\Re(A_{f})]^{2}+[\Im(A_{f})]^{2}$
of state $\left|f\right>$ reflects the degree of two-photon absorption.

This perturbative description allows a transformation into a frequency-domain picture.
Within the frequency-domain framework, the spectral field of the pulse
$E(\omega) \equiv \left|E(\omega)\right| \exp \left[ i\Phi(\omega)
\right]$ is given as the Fourier transform of $\varepsilon(t)$,
with $\left|E(\omega)\right|$ and $\Phi(\omega)$ being, respectively, the spectral
amplitude and phase of frequency $\omega$.
For the unshaped transform-limited (TL) pulse, $\Phi(\omega)=0$ for any $\omega$.
We also define the normalized spectral field $\widetilde{E}(\omega)
\equiv E(\omega) / \left|E_{0}\right| \equiv
\left|\widetilde{E}(\omega)\right| \exp \left[ i\Phi(\omega)
\right]$ that represents the pulse shape, where $\left|E_{0}\right|$ is the peak spectral amplitude.
This allows to clearly distinguish in the expressions given below between the dependence on the pulse intensity
and the dependence on the pulse shape.
The maximal spectral intensity $I_{0}$ is proportional to
$|E_{0}|^{2}$ ($I_{0}\propto|E_{0}|^{2}$). Different values of $I_{0}$
correspond to different temporal peak intensities
$I_{\scriptsize{\textrm{TL}}}$ of the transform-limited (TL) pulse.

As shown before for the weak-field regime \cite{silberberg_2ph_nonres1_2},
the 2$^{nd}$-order amplitude $A_{f}^{(2)}$ is given by
\begin{eqnarray}
A_{f}^{(2)} & = &
- \frac{1}{i \hbar^{2}} \left|E_{0}\right|^{2} A^{(2)}(\omega_{fg}) \; ,
\label{eq5:tot-amp-2nd-order} \\
A^{(2)}(\Omega) & = & \mu_{fg}^{2}
\int_{-\infty}^{\infty}\widetilde{E}(\omega)\widetilde{E}(\Omega-\omega)d\omega
\; , \label{eq6:expl-amp-2nd-order-w_fg}
\end{eqnarray}
where $\omega_{fg}$ is the $\left|g\right>$-$\left|f\right>$ transition frequency
and $\mu_{fg}^{2}$ is the corresponding real effective non-resonant two-photon coupling.
It is given by
$\mu_{fg}^{2} \equiv \sum_{n}\frac{\mu_{fn}\mu_{ng}}{\omega_{ng}-\omega_{0}}$, 
with $\omega_{0}$ being the carrier frequency of the pulse. 
Eqs.~(\ref{eq5:tot-amp-2nd-order})-(\ref{eq6:expl-amp-2nd-order-w_fg}) reflect the fact that
$A_{f}^{(2)}$ coherently interferes all the non-resonant two-photon
pathways from $\left|g\right>$ to $\left|f\right>$ of any combination of two absorbed photons with
frequencies $\omega$ and $\omega' = \omega_{fg} - \omega$.
Several such two-photon pathways are shown schematically in Fig.~\ref{fig_1}. %
The phase associated with each two-photon pathway is $\Phi(\omega) + \Phi(\omega_{fg}-\omega)$.
So, with the TL pulse all these pathways acquire zero phase and thus
interfere one with the other in a fully constructive way. With a
given spectrum $\left|E(\omega)\right|$, this leads to the maximal
$\left|A^{(2)}_{f}\right|$ and the maximal weak-field non-resonant
two-photon absorption.
%

The 4$^{th}$-order amplitude term $A_{f}^{(4)}$ is much more
complicated than $A_{f}^{(2)}$ and we have 
calculated it to be given by
\begin{eqnarray}
A_{f}^{(4)} & = & - \frac{1}{i \hbar^{4}} \left|E_{0}\right|^{4} \left[ A_{f}^{(4)\textrm{on-res}} + A_{f}^{(4)\textrm{near-res}} \right] \; ,
\label{eq7.1:amp-4th-order} \\
A_{f}^{(4)\textrm{on-res}} & = & i \pi A^{(2)}(\omega_{fg}) A^{(R)}(0) \; ,
\label{eq7.2:amp-4th-order} \\
A_{f}^{(4)\textrm{near-res}} & = & - \wp \int_{-\infty}^{\infty} d\delta \frac{1}{\delta} A^{(2)}(\omega_{fg}-\delta) A^{(R)}(\delta) \; ,
\label{eq7.3:amp-4th-order}
\end{eqnarray}
where $A^{(2)}(\Omega)$ is defined in Eq.~(\ref{eq6:expl-amp-2nd-order-w_fg}) and
\begin{eqnarray}
A^{(R)}(\Delta\Omega) & = & A^{(\textrm{non-res}R)}(\Delta\Omega) + A^{(\textrm{res}R)}(\Delta\Omega) \; ,
\label{eq8:amp-4th-order}\\
A^{(\textrm{non-res}R)}(\Delta\Omega) & = & (\mu_{ff}^{2}+\mu_{gg}^{2}) \int_{-\infty}^{\infty} \widetilde{E}(\omega + \Delta\Omega)
\widetilde{E}^{*}(\omega) d\omega  \; ,
\label{eq9:amp-4th-order} \\
A^{(\textrm{res}R)}(\Delta\Omega) & = & |\mu_{fn_{r}}|^{2} \left[ i \pi
\widetilde{E}(\omega_{fn_{r}}+\Delta\Omega) \widetilde{E}^{*}(\omega_{fn_{r}}) \right. \hspace{5cm} \nonumber \\  & & \left. 
         - \wp \int_{-\infty}^{+\infty} d\delta' \frac{1}{\delta'} \widetilde{E}(\omega_{fn_{r}}+\Delta\Omega-\delta')
                                                                   \widetilde{E}^{*}(\omega_{fn_{r}}-\delta') \right]. \hspace{0.7cm}  
\label{eq10:amp-4th-order}
\end{eqnarray}
%
This set of equations reflects the fact that $A_{f}^{(4)}$
interferes all the four-photon pathways from $\left|g\right>$ to
$\left|f\right>$ of any combination of three absorbed photons and
one emitted photon.
Several typical four-photon pathways are shown schematically in Fig.~\ref{fig_1}.

Each four-photon pathway can actually be divided into two two-photon
parts: (i) a non-resonant transition of two absorbed photons
$\omega$ and $\omega'$ with a frequency sum of $\Omega =
\omega + \omega' = \omega_{fg} - \delta$, and (ii) a Raman transition
of two photons $\omega_{R}$ and $\omega'_{R}$ with a frequency
difference of $\Delta\Omega = \omega_{R} - \omega'_{R}$. %
The border line between these parts is detuned by $\delta$ from either
$\left|f\right>$ or $\left|g\right>$ according to whether,
respectively, part (i) or part (ii)   
comes first (see Fig.~\ref{fig_1}).
The $A_{f}^{(4)\textrm{on-res}}$ and $A_{f}^{(4)\textrm{near-res}}$ terms of $A_{f}^{(4)}$
[Eqs.~(\ref{eq7.1:amp-4th-order})-(\ref{eq7.3:amp-4th-order})] interfere, respectively,
these on-resonant ($\delta = 0$) and near-resonant ($\delta \ne 0$) four-photon pathways.
The on-resonant pathways are excluded from $A_{f}^{(4)\textrm{near-res}}$ by the Cauchy's principle value operator $\wp$.
The integration taking place in these terms over the corresponding pathways   
is expressed using the product of two 
parameterized amplitudes, $A^{(2)}(\Omega)$ and
$A^{(R)}(\Delta\Omega)$, which originate from the different two-photon parts of the four-photon pathways.
The amplitude $A^{(2)}(\Omega)$ interferes all the non-resonant two-photon transitions     
[parts (i) above] with transition frequency $\Omega$, while
the amplitude $A^{(R)}(\Delta\Omega)$ interferes all the Raman transitions [parts (ii) above]
with transition frequency $\Delta\Omega$.
As given by Eqs.~(\ref{eq8:amp-4th-order})-(\ref{eq10:amp-4th-order}), $A^{(R)}(\Delta\Omega)$ includes two components.
The first component is $A^{(\textrm{non-res}R)}$ interfering all the Raman transitions
that are of non-resonant nature,
with $\mu^{2}_{gg}$ and $\mu^{2}_{ff}$ being the 
$\left|g\right>$-$\left|g\right>$ and
$\left|f\right>$-$\left|f\right>$ real effective non-resonant Raman
couplings due to all the non-resonantly coupled states $\left|n\right>$.
They are given by $\mu_{gg}^{2}=\sum_{n \ne n_{r}}
\left|\mu_{ng}\right|^{2} \left[\frac{1}{{\omega_{ng}-\omega_{0}}} +
\frac{1}{{\omega_{ng}+\omega_{0}}}\right]$ and $\mu_{ff}^{2}=\sum_{n
\ne n_{r}} \left|\mu_{nf}\right|^{2}
\left[\frac{1}{{\omega_{nf}-\omega_{0}}} +
\frac{1}{{\omega_{nf}+\omega_{0}}}\right]$.
Hence, their sum $(\mu_{gg}^{2} + \mu_{ff}^{2})$ [appearing in Eq.~(\ref{eq9:amp-4th-order})] is either positive or negative,
depending on the physical system and pulse spectrum.
The second component of $A^{(R)}(\Delta\Omega)$ is $A^{(\textrm{res}R)}$ interfering all the Raman transitions
that are of resonance-mediated nature via $\left|n_{r}\right>$,
with $\delta'$ being the detuning from $\left|n_{r}\right>$ (see Fig.~\ref{fig_1}).
These on-resonant ($\delta'=0$) and near-resonant ($\delta'\ne0$) transitions are interfered separately, respectively,
in the first and second terms of $A^{(\textrm{res}R)}$ [Eq.~(\ref{eq10:amp-4th-order})].
The Cauchy's principle value operator $\wp$ excludes the on-resonant transitions from the second term.


For a given physical system and a given pulse shape
$\widetilde{E}(\omega)$, a non-zero $A_{f}^{(2)}$ is proportional to
$|E_{0}|^{2}$ while a non-zero $A_{f}^{(4)}$ is proportional to
$|E_{0}|^{4}$, i.e., their ratio is proportional to $|E_{0}|^{2}$
or, equivalently, to $I_{0}$.
For a given $I_{0}$, the relative magnitude and relative sign between the real parts
$\Re[A_{f}^{(2)}]$ and $\Re[A_{f}^{(4)}]$ are generally determined by the pulse shape $\widetilde{E}(\omega)$ and
by the magnitudes and signs of the different Raman couplings ($\mu_{gg}^{2}$, $\mu_{ff}^{2}$, and $|\mu_{fn_{r}}|^{2}$).
The same applies also for the imaginary parts $\Im[A_{f}^{(2)}]$ and $\Im[A_{f}^{(4)}]$.

In the present work, for a set of intensities $I_{0}$,
the final $\left|f\right>$ population $P_{f}$ (i.e., the degree of two photon absorption)
is controlled via the pulse shape $\widetilde{E}(\omega)$.
Three different spectra $\left|\widetilde{E}(\omega)\right|$ are being considered,
with the control knobs being the various spectral phases $\Phi(\omega)$.
The three spectra are chosen such that 
their central spectral frequency $\omega_{0}$ is of 
no detuning, blue detuning, or red detuning from $\omega_{fg}/2$,
with $\left|\widetilde{E}(\omega_{fg}/2)\right| \simeq 0.5$ for the
detuned cases.
This set of spectra generally corresponds to the typical case, where
$A^{(4)}$ is negligible relative to $A^{(2)}$ in the weak-field limit
and becomes comparable to $A^{(2)}$ in the upper intermediate-field limit.
These spectral shifts do not change the sign of the various Raman couplings and
hardly change their magnitude. They do however affect the relative amplitude associated with the
different interfering pathways and, thus, their overall interference result.
For example, as shown below, in the cases studied here the spectral change from red- to blue-detuning leads to a change
in the relative sign between $\Im[A_{f}^{(2)}]$ and $\Im[A_{f}^{(4)}]$, and thus to a change in the nature of their interference
from destructive to constructive.

In general,
the dynamics and interference mechanisms discussed in this work
involve intra-term as well as (intensity-dependent) inter-term interferences
involving the multiphoton pathways of the
2$^{nd}$- and 4$^{th}$-order perturbative terms.


\section{\label{sec:results}Results}

The physical model system of the study is the sodium (Na) atom \cite{NIST}.
It includes the $3s$ ground state as $\left|g\right>$,
the $4s$ state as $\left|f\right>$,
the manifold of $p$ states as the $\left|n\right>$ manifold,
and the $7p$ state as $\left|n_{r}\right>$.
So, $A_{f}^{(2)} \equiv A_{4s}^{(2)}$ and $A_{f}^{(4)} \equiv A_{4s}^{(4)}$.
The transition frequency $\omega_{fg} \equiv \omega_{4s,3s} = 25740$~cm$^{-1}$
corresponds to two 777-nm photons and the transition frequency
$\omega_{fn_{r}} \equiv \omega_{7p,4s} = 12801$~cm$^{-1}$ corresponds to one 781.2-nm photon.
The atomic sodium is irradiated with phase-shaped linearly-polarized femtosecond pulses
having an intensity spectrum of $\sim$5-nm FWHM-bandwidth ($\sim$180~fs TL duration) centered around
a wavelength $\lambda_{0}$ tunable between 773 to 780~nm.
As a test case, for three different values of the central wavelength,
the present study uses the family of shaped pulses having $\pi$-step spectral phase patterns.
In the weak-file regime \cite{silberberg_2ph_nonres1_2} this family allows high
degree of control over the full accessible range of the two-photon absorption,
from zero to the maximal level (induced by the TL pulse).
Each such pattern is characterized by the $\pi$-step position $\omega_{step}$, with
$\Phi(\omega) \le \omega_{step}) = -\pi/2$ and $\Phi(\omega > \omega_{step}) = \pi/2$.

The understanding and analysis of the intermediate-field control
mechanism is conducted below using the frequency-domain picture
given in Eqs.~(\ref{eq4:time-amp})-(\ref{eq10:amp-4th-order}) and
corresponding numerical results for the Na system.
However, prior to the analysis, the extended perturbative picture and results
are validated by a comparison to exact non-perturbative results
that their own validity is confirmed first by a comparison to experiment.
The exact non-perturbative results have been calculated by the numerical propagation of the
time-dependent Schr\"{o}dinger equation equation (TDSE) using the fourth-order Runge-Kutta method.
The theoretically considered manifold of $p$-states is
from $3p$ to $8p$, including all the (1/2 and 3/2) fine-structure states \cite{NIST}.

\subsection{Non-perturbative calculations vs. experiment}

Experimentally, atomic sodium vapor is produced in a static chamber at
$300^{o}$C (Na partial pressure of $\sim$0.1 Torr) with 10-Torr Ar buffer gas.
It is irradiated at a 1-kHz repetition rate with shaped femtosecond laser pulses of three
different central spectral wavelengths: $\lambda_{0}$=773, 777, and 779.5~nm.
The corresponding spectral intensity bandwidth (FWHM) is, respectively, 5.5, 4.5, and 5~nm.
The 773-nm and 777-nm spectra are Gaussians,
while the 779.5-nm spectrum is a modified Gaussian having 
a slight asymmetry toward short wavelengths.
The slight change in the experimental spectral shape between the three cases
results from technical limitations and
is of no significance to the analysis and discussion presented below.
The laser pulses undergo shaping in a 4$f$ optical setup incorporating a pixelated
liquid-crystal spatial light phase modulator \cite{pulse_shaping}.
The effective spectral shaping resolution is $\delta\omega_{shaping}=2.05$~cm$^{-1}$ (0.125~nm) per pixel.
The experiment is conducted with different pulse energies.
Upon focusing, the corresponding temporal peak intensity of the transform-limited (TL) pulse at the
peak of the spatial beam profile $I_{\scriptsize{\textrm{TL}}}^{\scriptsize{\textrm{(profile-peak)}}}$
ranges from 5$\times$10$^{8}$ to 7$\times$10$^{10}$~W/cm$^2$.
Following the interaction with a pulse, the Na population
excited to the $4s$ state radiatively decays to the lower $3p$
state. The fluorescence emitted in the decay of the $3p$ state to
the $3s$ ground state serves as the relative measure for the
excited $4s$ population $P_{f} \equiv P_{4s}$. 
It is optically measured at 90$^{\circ}$ to the beam propagation direction
using a spectrometer coupled to a time-gated camera system.
The measured signal results from an integration over the spatial beam profile.

\begin{figure} 
\hspace*{-1.5cm}
\includegraphics[width=20cm]{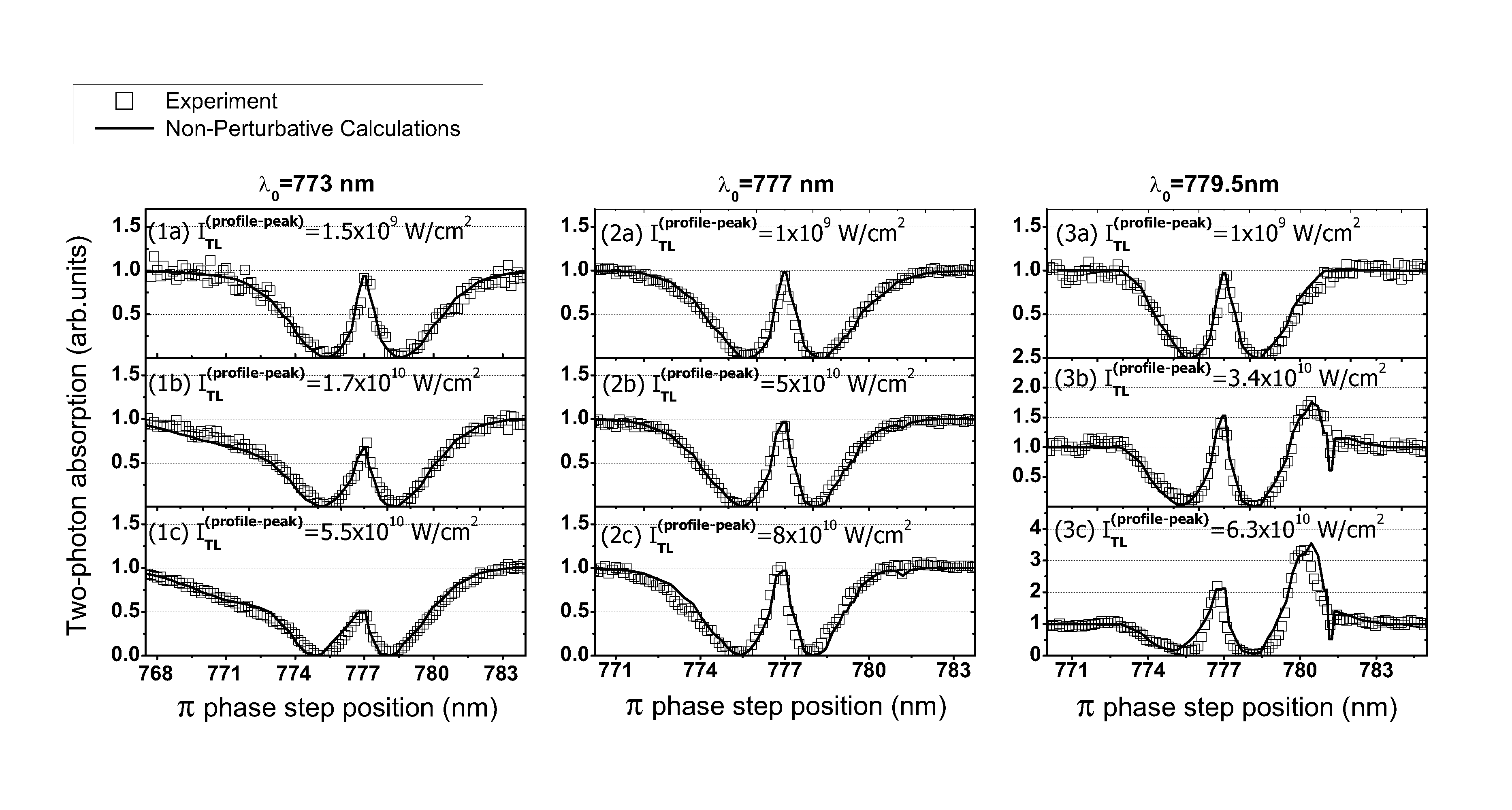}     
\caption{Experimental results (squares) and non-perturbative theoretical results (solid lines) 
for the two-photon absorption in Na induced by the shaped pulses having a $\pi$-step spectral phase pattern.
The results include an integration over the experimental spatial beam profile.
The traces show the final $4s$ population $P_{4s}$ as a function of the step position $\omega_{step}$.
The value of $P_{4s}$ is normalized by the final population $P_{4s,\textrm{TL}}$ excited by the corresponding transform-limited (TL) pulse.
Each column 
corresponds to a different central spectral wavelength  
$\lambda_{0}$=773, 777, and 779.5~nm with a different pulse energy in each row (increasing from top to bottom).
Each pulse energy corresponds to a different transform-limited intensity at the peak of the
spatial beam profile $I_{\scriptsize{\textrm{TL}}}^{\scriptsize{\textrm{(profile-peak)}}}$.
} \label{fig_2}
\end{figure}

Figure~\ref{fig_2} compares the non-perturbative theoretical results (solid lines)
with the experimental results (squares) for the two-photon absorption in Na.
Each column in the figure corresponds to a different spectral case
of $\lambda_{0}$=773, 777, or 779.5~nm with different pulse
energies, i.e., different
$I_{\scriptsize{\textrm{TL}}}^{\scriptsize{\textrm{(profile-peak)}}}$.
The pulse energy increases from top to bottom within a single
column.
The traces show the final $4s$ population $P_{4s}$ as a function of the $\pi$ phase step position $\omega_{step}$.
Each of the traces is normalized by the final $4s$ population $P_{4s,\textrm{TL}}$ excited by the corresponding TL pulse.
The weak-field $\pi$-traces are given in the first-row panels [(a)-panels] of Fig.~\ref{fig_2}.
The non-perturbative theoretical results shown in Fig.~\ref{fig_2} account for the experimental
integration over the spatial beam profile. Each of the presented traces results from an appropriately-weighted
integration over a set of calculations conducted each with
a different single value of $I_{0}$. 

As can be seen, there is an excellent agreement between the experimental results ("real experiment")
and the non-perturbative results ("computer experiment").
Hence, the accuracy of the latter is confirmed for the present intermediate-field excitation of Na.

\subsection{Perturbative calculations vs. non-perturbative calculations}

Next, the confirmed non-perturbative calculations are used to
validate the intermediate-field perturbative results and to identify
the intensity limit of the intermediate-field regime for the present
Na excitation, i.e., the intensity up to which no perturbative order
beyond the 4$^{th}$ one is needed to be included.
From this point on, the analysis and discussion are conducted based on the theoretical results.
So, for completeness, the three intensity spectra considered theoretically are of a perfect
Gaussian shape with 5-nm bandwidth (FWHM) centered around $\lambda_{0}$=773.5, 777, and 780~nm.
As mentioned above, these $\lambda_{0}$ values correspond to
blue detuning, no detuning, and red detuning of $\omega_{0}$ from $\omega_{4s,3s}/2$, with
$\left|\widetilde{E}(\omega_{4s,3s})\right| \simeq 0.5$ for the detuned cases. 

Figure~\ref{fig_3} compares the theoretical non-perturbative results (thick gray lines)
and perturbative results calculated numerically using Eqs.~(\ref{eq4:time-amp})-(\ref{eq10:amp-4th-order}) (thick black lines).
The perturbative calculations use a frequency grid with a bin size equal to the
experimental spectral shaping resolution $\delta\omega_{shaping}$.
The presented $\pi$-traces are given on a TL-normalized scale of $P_{4s}/P_{4s,\textrm{TL}}$ (left-hand y-axis)
as well as on an absolute scale of $P_{4s}$ (right-hand y-axis).
The x-axis is the $\pi$ phase step position $\omega_{step}$.
Also here, each column corresponds to a different spectral case of $\lambda_{0}$ with
different (single-valued) intensities $I_{0}$, i.e., different $I_{\scriptsize{\textrm{TL}}}$.
The $I_{0}$ increases from top to bottom within a single column.
Shown are several examples out of the full set of results.
The weak-field traces are given in the first-row panels [(a)-panels] of Fig.~\ref{fig_3}.
Actually, on a TL-normalized scale they are all identical and independent of $\lambda_{0}$ \cite{silberberg_2ph_nonres1_2}.
For comparison, the TL-normalized weak-field trace is also given in all the other panels (thin black line).
As can be seen from the figure, the perturbative results reproduce
the exact non-perturbative results up to the $I_{0}$ that corresponds
to a TL peak intensity of $I_{\scriptsize{\textrm{TL}}}= 2.5 \times 10^{10}$~W/cm$^{2}$
[(c)-panels of Fig.~\ref{fig_3}].
This is the intensity limit of the 4$^{th}$-order intermediate-field regime for the present Na excitation.
As the last-row panels [(d)-panels] of Fig.~\ref{fig_3} show, 
the 4$^{th}$-order perturbative description is not sufficient above this intensity.
The corresponding intensity limit of the weak-field regime, where $A_{f} \approx A_{f}^{(2)}$,
is $I_{\scriptsize{\textrm{TL}}} \approx 5 \times 10^{8}$~W/cm$^{2}$.

\subsection{Intermediate-field features of the $\pi$-trace}
\label{subsec3:results}

\begin{figure} 
\hspace*{-2cm}\includegraphics[width=20.5cm]{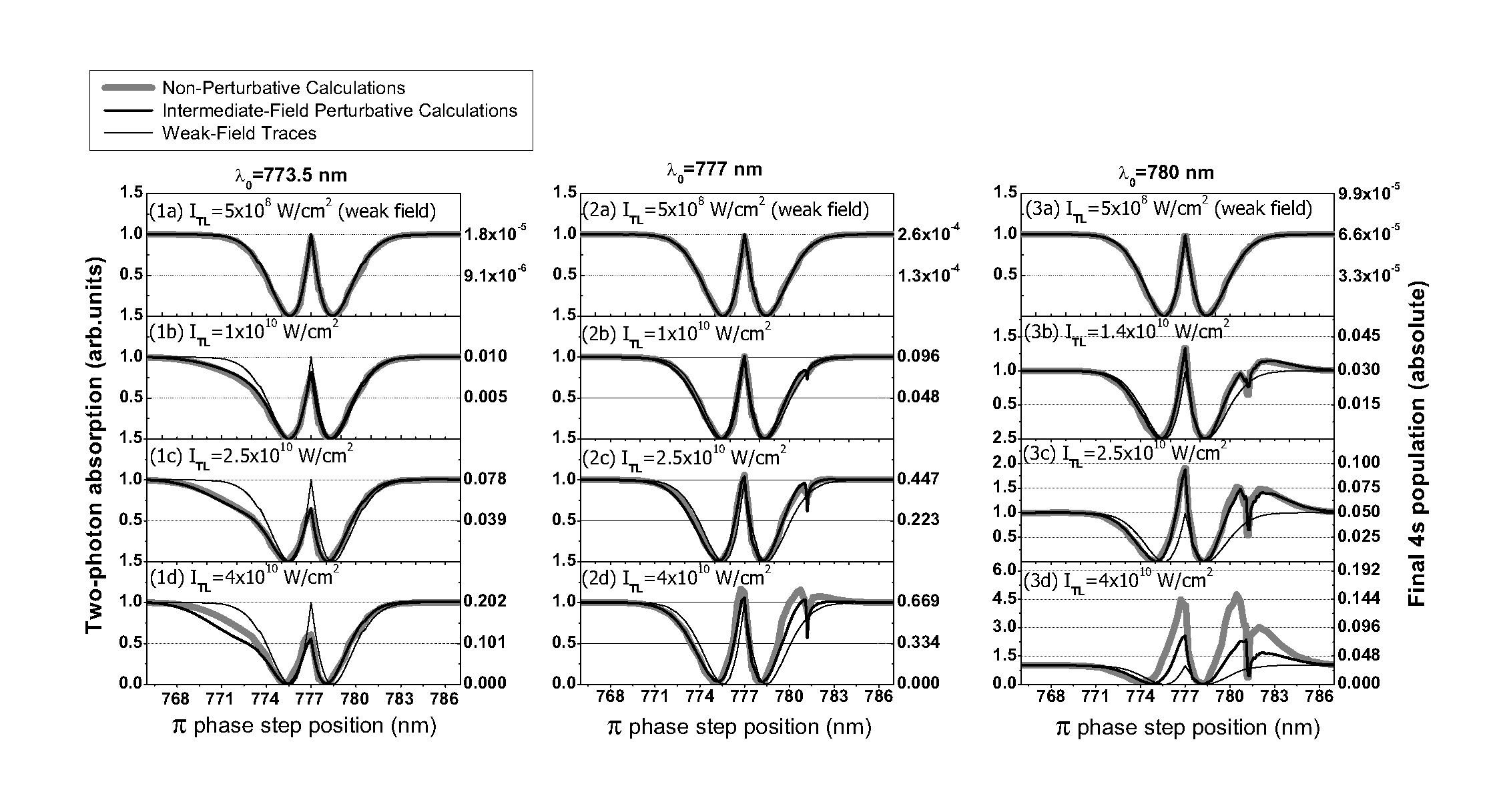}     
\caption{Non-perturbative theoretical results (gray thick lines)
and 4$^{th}$-order perturbative theoretical results (black thick lines)
for the two-photon absorption in Na induced by the shaped pulses having a $\pi$-step spectral phase pattern.
As a 4$^{th}$-order calculations, the perturbative calculations include contributions from the 2$^{nd}$ and 4$^{th}$ orders.
The traces show the final $4s$ population $P_{4s}$ as a function of the step position $\omega_{step}$.
The right-hand y-axis scale is the absolute value of $P_{4s}$.
The left-hand y-axis scale corresponds to $P_{4s}$ normalized by the final population $P_{4s,\textrm{TL}}$
excited by the corresponding transform-limited (TL) pulse.
Each column 
corresponds to a different central spectral wavelength  
$\lambda_{0}$=773.5, 777, and 780~nm with a different (single-valued) spectral intensity $I_{0}$ in each row (increasing from top to bottom).
Each $I_{0}$ corresponds to a different transform-limited peak intensity $I_{\scriptsize{\textrm{TL}}}$.
The weak-field traces are given in the first-row panels [(a)-panels].
For comparison, they are also shown in all the other panels (black thin lines) on the TL-normalized scale.
} \label{fig_3}
\end{figure}

The prominent features of the intermediate-field $\pi$-traces are presented below.
As previously shown \cite{silberberg_2ph_nonres1_2} and can be seen in Fig.~\ref{fig_3} [the (a)-panels],
the weak-field TL-normalized $\pi$-trace is symmetric around
$\omega_{step} = \omega_{4s,3s}/2$ (777~nm) and 
is identical for any $\omega_{0}$ ($\lambda_{0}$). Its shape is determined only by the spectral bandwidth of the pulse.
Also, since it is TL-normalized, it is independent of the intensity $I_{0}$ [see Eq.~(\ref{eq5:tot-amp-2nd-order})].
However, when deviating from the weak-field regime with the 4$^{th}$ perturbative order playing a role,
the TL-normalized $\pi$-trace losses its weak-field symmetry and its shape 
becomes dependent on both $\omega_{0}$ and $I_{0}$.
At a given $\omega_{0}$, the degree of deviation from the weak-field shape depends on $I_{0}$.

As mentioned above, in the weak-field regime the maximal non-resonant two-photon absorption 
is induced by the TL pulse. 
Additionally, as can be seen in the figure,
the same maximal weak-field two-photon absorption is also induced
by the shaped pulse with $\omega_{step} = \omega_{4s,3s}/2$ (777~nm),
i.e. $P_{4s,\textrm{step@}\omega_{4s,3s}/2} = P_{4s,\textrm{TL}}$ or 
$R_{\textrm{step@}\omega_{4s,3s}/2} \equiv P_{4s,\textrm{step@}\omega_{4s,3s}/2} / P_{4s,\textrm{TL}} = 1$.
Both pulses induce fully constructive interferences (i.e., zero relative phase)
between all the $\left|g\right>$-$\left|f\right>$ two-photon pathways.
Conversely,
in the intermediate-field regime a $\pi$-phase step at $\omega_{step} = \omega_{4s,3s}/2$
leads to a two-photon absorption that, in general, is different from the TL absorption. 
The difference is highly pronounced when $\omega_{0}$ is detuned
from $\omega_{4s,3s}/2$ [Fig.~\ref{fig_3} - columns(1) and (3)] and
is weakly pronounced when $\omega_{0} = \omega_{4s,3s}/2$
[Fig.~\ref{fig_3} - column (2)],
with $R_{4s,\textrm{step@}\omega_{4s,3s}/2} < 1$ for a blue
detuning, $R_{4s,\textrm{step@}\omega_{4s,3s}/2} \approx 1$ for no
detuning, and $R_{4s,\textrm{step@}\omega_{4s,3s}/2} > 1$ for a red
detuning.
For example, at the intermediate-field limit, with $I_{0}$ of
$I_{\scriptsize{\textrm{TL}}}= 2.5 \times 10^{10}$~W/cm$^{2}$,
$R_{\textrm{step@}\omega_{4s,3s}/2}$ reaches a value of 0.65, 1.15,
2.0 for $\lambda_{0}$=773.5~nm [Fig.~\ref{fig_3}(1c)],
$\lambda_{0}$=777~nm [Fig.~\ref{fig_3}(2c)], and
$\lambda_{0}$=780~nm [Fig.~\ref{fig_3}(3c)], respectively.
The corresponding degree of attenuation or enhancement over the TL absorption , i.e.,
$\left| R_{\textrm{step@}\omega_{4s,3s}/2} \right|$, increases as $I_{0}$ increases.
Also, as $I_{0}$ increases, this enhancement/attenuation effect occurs (with variable corresponding values)
over an increased range of $\omega_{step}$ around $\omega_{4s,3s}/2$.

In the spectrally detuned cases of $\omega_{0} \ne \omega_{4s,3s}/2$ with $\left|\widetilde{E}(\omega_{4s,3s}/2)\right| \simeq 0.5$
(as is the case here; see above)
a $\pi$ phase step that is positioned around $\omega_{0}$, i.e., $\omega_{step} \approx \omega_{0}$,
leads to intermediate-field values of the TL-normalized trace [Fig.~\ref{fig_3} - columns (1) and (3), thick solid lines]
that are systematically either attenuated or enhanced relative to
the weak-field trace [Fig.~\ref{fig_3} - columns (1) and (3), thin solid lines].
The attenuation occurs for blue-detuned $\lambda_{0}$ and the enhancement occurs for red-detuned $\lambda_{0}$.
Also here, the degree of attenuation or enhancement relative to the
weak-field case at a given $\omega_{step}$ increases as $I_{0}$
increases. In the red-detuned case, at high enough intensity
$I_{0}$, the corresponding two-photon absorption even exceeds the TL
absorption [see Fig.~\ref{fig_3}(3c)].

An additional intermediate-field feature is the dip observed when $\omega_{step} = \omega_{7p,4s}$ (781.2~nm),
which does not exist in the weak-field trace. It occurs when the field amplitude
$\left|\widetilde{E}(\omega_{fn_{r}})\right| \equiv \left|\widetilde{E}(\omega_{7p,4s})\right|$ at $\omega_{7p,4s}$
significantly deviates from zero.
Hence, this 781.2-nm dip is very prominent for $\lambda_{0}$=780~nm [Fig.~\ref{fig_3} - column (3)],
weakly noticeable for $\lambda_{0}$=777~nm [Fig.~\ref{fig_3} - column (2)],
and hardly noticeable for $\lambda_{0}$=773.5~nm [Fig.~\ref{fig_3} - column (1)].

Lastly, it is important to mention that within the
intermediate-field regime, due to the 4$^{th}$-order term
$A_{4s}^{(4)}$, the absolute $4s$ population $P_{4s,\textrm{TL}}$
excited by a given shaped pulse strongly deviates from the
weak-field intensity dependence of $(I_{0})^{2}$ in all the
$\lambda_{0}$ cases. Specifically, it applies also to the
non-detuned case of $\lambda_{0}$=777~nm, even though the
corresponding intermediate-field shape of the TL-normalized
$\pi$-trace does not deviate much from the weak-field shape. This
can be seen in Fig.~\ref{fig_3}, for example, in the absolute
population values that are induced by the TL pulse (i.e., the
asymptotes of the traces) at different intensities.


\section{Discussion}
\label{sec:discussion}

\subsection{General considerations}

The discussion below analyzes the coherent source for the intermediate-field features described above,
using the frequency-domain picture given in Eqs.~(\ref{eq4:time-amp})-(\ref{eq10:amp-4th-order}).
The different intermediate-field features,
i.e., the deviation of the intermediate-field trace from the weak-field trace,
originate from the different dependence of the 2$^{nd}$- and 4$^{th}$-order amplitudes,
$A_{4s}^{(2)}$ and $A_{4s}^{(4)}$, on the pulse shape.

For a given spectrum $\left|E(\omega)\right|$,
the on-resonant term $A_{4s}^{(4)\textrm{on-res}}$ of $A_{4s}^{(4)}$ [Eq.~(\ref{eq7.2:amp-4th-order})] is proportional to $A_{4s}^{(2)}$.
It is so since $A^{(R)}(\Delta\Omega=0)$ depends only on the spectral intensity:
all the corresponding Raman transitions involve two identical photons and their
amplitudes in Eqs.~(\ref{eq8:amp-4th-order})-(\ref{eq10:amp-4th-order})
are of the form $\widetilde{E}(\omega)\widetilde{E}^{*}(\omega) = \left|\widetilde{E}(\omega)\right|^{2}$.
Thus, the difference in the phase dependence of $A_{4s}^{(2)}$ and $A_{4s}^{(4)}$ originates only from the
near-resonant term $A_{f}^{(4)\textrm{near-res}}$ of $A_{4s}^{(4)}$,  
which interferes all the near-resonant four-photon pathways of non-zero detunings $\delta \ne 0$
from $3s$ or $4s$ (see above).

As Eq.~(\ref{eq7.3:amp-4th-order}) shows, $A_{f}^{(4)\textrm{near-res}}$ is given by a proper
$\wp$-integration over all the non-zero $\delta$ values. Its phase dependence
originates from the phase dependence of the integrands $A^{(2)}(\omega_{4s,3s}-\delta)$ and $A^{(R)}(\delta)$.
Their dependence on the spectral phase pattern $\Phi(\omega)$ is reflected in their values for a given $\delta$ and, more importantly,
in their functional dependence on $\delta$.
So, the $\wp$-integration yields different results for different phase patterns.
Due to the $1/\delta$ weighting (and its sign change for $\pm$$\left|\delta\right|$),
the $\wp$-integration result is dominated by the integration over small values of $\left|\delta\right|$ and is highly sensitive 
to the degree of symmetry of the integrand $A^{(2)}(\omega_{4s,3s}-\delta) A^{(R)}(\delta)$ around $\delta=0$,
i.e., how different are its values for $\pm$$\left|\delta\right|$. 
Below, the intermediate-field two-photon absorption is analyzed based on the
$\delta$-dependence of $A^{(2)}(\omega_{4s,3s}-\delta)$ and $A^{(R)}(\delta)$
for several representative cases of the $\pi$-step position $\omega_{step}$ with the different $\lambda_{0}$.

The analysis of the amplitude $A^{(R)}(\delta)$,
which is contributed by all the Raman parts of the $\delta$-detuned four-photon pathways,
is simplified 
by including its component $A^{(\textrm{res}R)}(\delta)$ [see Eq.~(\ref{eq8:amp-4th-order})]
only for the study of the dip feature at $\omega_{step}=\omega_{7p,4s}$ (781.2~nm).
As described above, $A^{(\textrm{res}R)}(\delta)$ interferes those Raman parts that are resonance-mediated via the $7p$ state.
This line of analysis is supported by the discussion given below and
by the perturbative results presented in Fig.~\ref{fig_4} for the $\pi$-traces
at $I_{0}$ of $I_{\scriptsize{\textrm{TL}}}= 2.5 \times 10^{10}$~W/cm$^{2}$ (i.e., the intermediate-field limit).
In addition to the real TL-normalized $\pi$-traces (gray thick lines),
which are also shown in the (c)-panels of Fig.~\ref{fig_3},
the figure displays the $\pi$-traces (black thick lines) calculated with 
$A^{(R)}(\delta) = A^{(\textrm{non-res}R)}(\delta)$,
i.e., with artificially setting $A^{(\textrm{res}R)}(\delta)$ to zero for any $\delta$.
As can be seen, the resonance-mediated Raman transitions via $7p$ are of 
significance only for discussing the dip at $\omega_{step}=\omega_{7p,4s}$ (781.2~nm).
For completeness, the figure also displays the weak-field trace (black thin lines; shown also in Fig.~\ref{fig_3})
that originates only from $A_{f}^{(2)}$ and, thus, is not affected by changes in the Raman part.

The non-resonant Raman term $A^{(\textrm{non-res}R)}(\delta)$ [Eq.~(\ref{eq9:amp-4th-order})]
coherently integrates all the non-resonant Raman amplitudes $\widetilde{E}(\omega+\delta)\widetilde{E}^{*}(\omega)$
contributed by all the possible pairs of photons
with a frequency difference of $\delta$ ($\omega$ is scanned across the spectrum).
With $\Phi(\omega)=0$ (TL pulse) or any $\pi$ spectral phase step, the amplitude contributed by any such pair of photons
is a real (positive or negative) quantity and, thus, so is the resulting $A^{(\textrm{non-res}R)}(\delta)$.
So, with these phase patterns, the relation
$A^{(\textrm{non-res}R)}($$-$$\left|\delta\right|) =
A^{(\textrm{non-res}R)}($$+$$\left|\delta\right|)$ holds for any
$\delta$, i.e., $A^{(\textrm{non-res}R)}(\delta)$ is symmetric
around $\delta=0$.
This symmetry can be seen in Fig.~\ref{fig_5} that shows, as an example,
the value of $A^{(\textrm{non-res}R)}(\delta)$ (gray thick lines)
as a function of $\delta$ for the TL pulse and for the shaped pulses
with $\omega_{step}=\omega_{4s,3s}/2$ and $\omega_{step}=\omega_{0}$.
Each column corresponds to a different $\lambda_{0}$ ($\omega_{0}$).
The x-axis values are actually the normalized detuning values $\delta/\Delta\omega$, with $\Delta\omega$ being the
bandwidth of the intensity spectrum. 
As can also be seen in the figure,
$A^{(\textrm{non-res}R)}(\delta)$ is maximal at $\delta=0$,
with a value that depends only on the spectral intensity $\left|\widetilde{E}(\omega)\right|^{2}$, i.e., is independent of $\Phi(\omega)$.

The other term $A^{(2)}(\omega_{4s,3s}-\delta)$ [Eq.~(\ref{eq6:expl-amp-2nd-order-w_fg})]
coherently integrates all the non-resonant two-photon transition amplitudes
$\widetilde{E}(\omega)\widetilde{E}(\omega_{4s,3s}-\delta)$
contributed by all the possible pairs of photons with a frequency sum of $\omega_{4s,3s}-\delta$.
Also here, with $\Phi(\omega)=0$ (TL pulse) or any $\pi$ spectral phase step,
the amplitude contributed by any such pair of photons
is a real (positive or negative) quantity and, thus, so is the resulting $A^{(2)}(\omega_{4s,3s}-\delta)$.
Figure~\ref{fig_5} displays, together with the Raman part data, 
the corresponding $A^{(2)}(\omega_{4s,3s}-\delta)$ (black thick lines) as a function of $\delta/\Delta\omega$
in the different cases.
The zone of small $\left|\delta\right|$ around $\delta=0$, which is the most contributing to $A_{f}^{(4)\textrm{near-res}}$ (see above),
is also indicated schematically. 
As can be seen from the figure, the $\delta$-dependence of $A^{(2)}(\omega_{4s,3s}-\delta)$ around $\delta=0$
is different from one case to the other and depends on both the spectral phase pattern and $\lambda_{0}$ ($\omega_{0}$).
Each panel also shows, for comparison, the $A^{(2)}(\omega_{4s,3s}-\delta)$ trace of the corresponding TL pulse
(black thin line).

\subsection{Intermediate-field two-photon absorption: shaped pulse with $\omega_{step} = \omega_{4s,3s}/2$ vs.~TL pulse}
\label{subsec2:discussion}

\begin{figure} 
\hspace*{-1.8cm}\includegraphics[width=19.7cm]{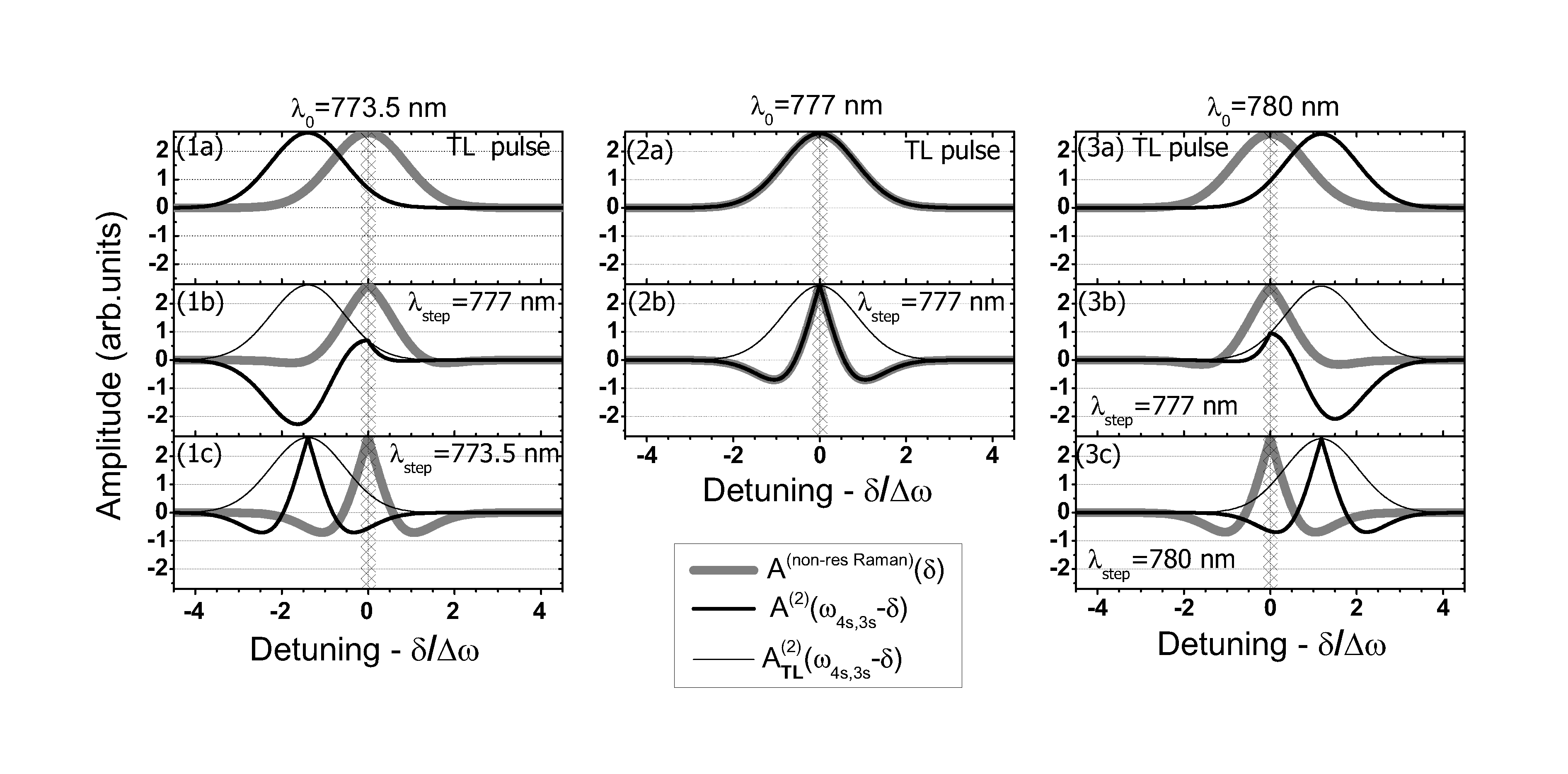}     
\vspace*{-0.7cm}
\caption{
Theoretical results for different quantities that are introduced by the frequency-domain 4$^{th}$-order perturbative description.
They are discussed in the text for explaining the different intermediate-field features.
Each panel shows the Raman term $A^{(\textrm{non-res}R)}(\delta)$ (gray thick lines)
and the two-photon transition term $A^{(2)}(\omega_{3s,3s}-\delta)$ (black thick lines)
as a function of the detuning $\delta$ (see text).
The detuning values (x-axis scale) are actually given as the normalized values $\delta/\Delta\omega$,
with $\Delta\omega$ being the bandwidth of the intensity spectrum of the pulse.
The zone of small $\left|\delta\right|$ around $\delta=0$, which is the most contributing to $A_{f}^{(4)\textrm{near-res}}$ (see text),
is indicated schematically (dashed area).
Each column corresponds to a different central spectral wavelength $\lambda_{0}$=773.5, 777, and 780~nm.
Each row corresponds to a different pulse shape: (a) transform-limited (TL) pulse,
(b) shaped pulse with $\omega_{step}=\omega_{4s,3s}/2$ (777~nm), and
(c) shaped pulse with $\omega_{step}=\omega_{0}$ (the corresponding central spectral frequency).
For comparison, each panel also shows the $A^{(2)}(\omega_{3s,3s}-\delta)$ term of the corresponding TL pulse (black thin lines).
\vspace{4cm}
} \label{fig_5}
\end{figure}

The first-row panels [(a)-panels] of Fig.~\ref{fig_5} correspond to
the TL pulse of the different $\lambda_{0}$ ($\omega_{0}$) cases.
Since the TL pulse induces fully constructive interferences among all the
two-photon pathways that contribute to $A^{(2)}(\Omega)$ for any $\Omega$, the
corresponding value $A_{\textrm{TL}}^{(2)}(\Omega=\omega_{4s,3s}-\delta)$
is actually the maximal one (positive and real) for any given $\delta$.
As can be seen from Eq.~(\ref{eq6:expl-amp-2nd-order-w_fg}),
$A_{\textrm{TL}}^{(2)}(\Omega=\omega_{4s,3s}-\delta)$ is actually a convolution of
the corresponding spectrum $\left|\widetilde{E}(\omega)\right|$.
Thus, with a Gaussian spectrum around $\omega_{0}$,
it is peaked at $\Omega_{\textrm{peak}} = 2 \omega_{0}$,
i.e., at $\delta_{\textrm{peak}} = \omega_{4s,3s} - 2 \omega_{0}$.
So, for the different cases of $\omega_{0}$ one obtains the following behavior:
(i) For $\omega_{0} > \omega_{4s,3s}/2$ [$\lambda_{0}=$773.5~nm, Fig.~\ref{fig_5}(1a)] --
$\delta_{\textrm{peak}} < 0$ and $A^{(2)}(\omega_{4s,3s}-\delta)$ monotonically decreases
around $\delta=0$ as $\delta$ increases from negative to positive values,
i.e., $A^{(2)}(\omega_{4s,3s}-\left|\delta\right|) > A^{(2)}(\omega_{4s,3s}+\left|\delta\right|)$ for small $\left|\delta\right|$;
(ii) For $\omega_{0} < \omega_{4s,3s}/2$ [$\lambda_{0}=$780~nm, Fig.~\ref{fig_5}(3a)] --
$\delta_{\textrm{peak}} > 0$ and $A^{(2)}(\omega_{4s,3s}-\delta)$ monotonically increases
around $\delta=0$ upon the negative-to-positive increase of $\delta$,
i.e., $A^{(2)}(\omega_{4s,3s}-\left|\delta\right|) < A^{(2)}(\omega_{4s,3s}+\left|\delta\right|)$ for small $\left|\delta\right|$;
(iii) For $\omega_{0} = \omega_{4s,3s}/2$ [$\lambda_{0}=$777~nm, Fig.~\ref{fig_5}(2a)] --
$\delta_{\textrm{peak}} = 0$ and $A^{(2)}(\omega_{4s,3s}-\delta)$ is symmetric around $\delta=0$,
i.e., $A^{(2)}(\omega_{4s,3s}-\left|\delta\right|) = A^{(2)}(\omega_{4s,3s}+\left|\delta\right|)$.

The second-row panels [(b)-panels] of Fig.~\ref{fig_5} correspond to the shaped pulse
with $\omega_{step} = \omega_{4s,3s}/2$ (777~nm).
Generally, when a $\pi$ phase step is positioned at $\omega_{step}$, the value of $A^{(2)}(\Omega=2 \omega_{step})$ is equal to
the corresponding TL value $A_{\textrm{TL}}^{(2)}(\Omega=2\omega_{step})$
due to fully constructive interferences among all the involving two-photon pathways.
As for the TL pulse, the phases associated with these pathways are all zero.
However, as $\Omega$ deviates from $2 \omega_{step}$, the value of $A^{(2)}(\Omega)$
gradually reduces with comparable magnitude for positive and negative deviations.
Hence, 
a $\pi$-step at $\omega_{step} = \omega_{4s,3s}/2$
yields a peak of $A^{(2)}(\omega_{4s,3s}-\delta)$ at $\delta=0$ (with the TL value).
When $\omega_{0} = \omega_{4s,3s}/2$ [$\lambda_{0}$=777~nm; Fig.~\ref{fig_5}(2b)], this peak is a global one
and $A^{(2)}(\omega_{4s,3s}-\delta)$ has a perfect symmetry around $\delta=0$,
i.e., $A^{(2)}(\omega_{4s,3s}-\left|\delta\right|) = A^{(2)}(\omega_{4s,3s}+\left|\delta\right|)$.
When $\omega_{0} \ne \omega_{4s,3s}/2$
[$\lambda_{0}=$773.5 and 780~nm; Fig.~\ref{fig_5}(1b) and (3b)],
this peak is a local one with only an approximate symmetry of $A^{(2)}(\omega_{4s,3s}-\delta)$  around $\delta = 0$,
i.e., $A^{(2)}(\omega_{4s,3s}-\left|\delta\right|) \approx A^{(2)}(\omega_{4s,3s}+\left|\delta\right|)$
for small $\left|\delta\right|$.

Based on the above analysis of $A^{(\textrm{non-res}R)}(\delta)$ and $A^{(2)}(\omega_{4s,3s}-\delta)$,
considering also the integrand factor $1/\delta$ that is anti-symmetric around $\delta=0$,
the magnitude of $A_{4s}^{(4)\textrm{near-res}}$ 
can be compared between the TL pulse case $\left[A_{4s,\textrm{TL}}^{(4)\textrm{near-res}}\right]$
and the shaped pulse case of $\omega_{step} = \omega_{4s,3s}/2$
$\left[A_{4s,\textrm{step@}\omega_{4s,3s}/2}^{(4)\textrm{near-res}}\right]$:
(i) In the no-detuning case of $\lambda_{0}$=777~nm --
$A_{4s,\textrm{TL}}^{(4)\textrm{near-res}} =
A_{4s,\textrm{step@}\omega_{4s,3s}/2}^{(4)\textrm{near-res}} = 0$,
since the amplitudes contributed to the $\wp$-integral by the
four-photon pathways of positive and negative detuning
$\pm$$\left|\delta\right|$ are equal and, thus, cancel out each
other;
(ii) In the red- and blue-detuning cases of $\lambda_{0}$=773.5 and 780~nm --
$\left|A_{4s,\textrm{TL}}^{(4)\textrm{near-res}}\right| >
 \left|A_{4s,\textrm{step@}\omega_{4s,3s}/2}^{(4)\textrm{near-res}}\right|$,
since the amplitudes contributed by the four-photon pathways of detunings $\pm$$\left|\delta\right|$
are approximately equal one to the other for the shaped pulse of $\omega_{step} = \omega_{4s,3s}/2$
while they are significantly different one from the other for the TL pulse.

As discussed below, important is also the sign of $A_{4s}^{(4)\textrm{near-res}}$ relative to $A_{4s}^{(2)}$,
which is determined by the pulse spectrum and by the non-resonant Raman couplings sum $(\mu_{3s,3s}^{2} + \mu_{4s,4s}^{2})$.
The sign of $A_{4s,\textrm{TL}}^{(4)\textrm{near-res}}$ relative to
$A_{4s,\textrm{TL}}^{(2)}$ and the sign of
$A_{4s,\textrm{step@}\omega_{4s,3s}/2}^{(4)\textrm{near-res}}$
relative to $A_{4s,\textrm{step@}\omega_{4s,3s}/2}^{(2)}$ are
obtained here to be the same for any $\lambda_{0}$.
However, upon a blue-to-red spectral shift of $\lambda_{0}$ (i.e., from 773.5 to 780~nm) these signs change from positive to negative,
with the sign flip occurring when passing via $\lambda_{0}$=777~nm.
The sign change results from a change in the relative magnitude of the amplitudes contributed by
the negatively-detuned and positively-detuned four-photon pathways,
while $(\mu_{3s,3s}^{2} + \mu_{4s,4s}^{2})$ keeps its sign and is effectively constant over the
whole $\lambda_{0}$ range considered here.

For comparing the two-photon absorption induced by the TL pulse vs.~the absorption induced by the shaped pulse
of $\omega_{step} = \omega_{4s,3s}/2$, one needs to consider the coherent amplitude addition of $A_{4s}^{(2)}$ and $A_{4s}^{(4)}$.
For the TL pulse and for any shaped pulse with a $\pi$-step phase pattern,
$A_{4s}^{(2)}$ [Eq.~(\ref{eq5:tot-amp-2nd-order})] is an imaginary quantity, 
i.e., $A_{4s}^{(2)} = \Im\left[A_{4s}^{(2)}\right]$,
while $A_{4s}^{(4)}$ is a complex quantity.
When only $A^{(\textrm{non-res}R)}$ is included in $A^{(R)}$ (see above),
the corresponding $A_{4s}^{(4)\textrm{near-res}}$ component is real
while the corresponding $A_{4s}^{(4)\textrm{on-res}}$ component is imaginary.
Due to the $1/i$ preceding factor of $A_{4s}^{(4)}$ [Eq.~(\ref{eq7.1:amp-4th-order})],
they contribute, respectively, solely to the imaginary and real parts of $A_{4s}^{(4)}$, i.e.,
$\Im\left[A_{4s}^{(4)}\right] \propto A_{4s}^{(4)\textrm{near-res}}$ and
$\Re\left[A_{4s}^{(4)}\right] \propto A_{4s}^{(4)\textrm{on-res}}$.
Thus, the interferences between the pathways groups of $A_{4s}^{(2)}$ and of $A_{4s}^{(4)}$
actually take place between those included in $A_{4s}^{(2)}$ and
those included in $\Im\left[A_{4s}^{(4)}\right]$, i.e., in $A_{4s}^{(4)\textrm{near-res}}$.

Since the following relations are satisfied here: (i)
$A_{4s,\textrm{step@}\omega_{4s,3s}/2}^{(2)} =
A_{4s,\textrm{TL}}^{(2)} \equiv A_{4s,*}^{(2)}$, (ii) $P_{4s} =
\left|A_{4s}\right|^{2} = \left|
A_{4s}^{(2)}+\Im\left[A_{4s}^{(4)}\right] \right|^{2} + \left|
\Re\left[A_{4s}^{(4)}\right] \right|^{2}$, (iii)
$\Re\left[A_{4s}^{(4)}\right] \propto i A_{4s}^{(2)}$ [see
Eq.~(\ref{eq7.2:amp-4th-order})], and (iv) $A_{4s}^{(2)} \propto
I_{0}$ while $A_{4s}^{(4)} \propto I_{0}^{2}$ [see
Eqs.~(\ref{eq5:tot-amp-2nd-order}) and (\ref{eq7.1:amp-4th-order})],
the TL-normalized $\pi$-trace value
$R_{\textrm{step@}\omega_{4s,3s}/2}$ (real and positive) corresponding to $\omega_{step} = \omega_{4s,3s}/2$ is given as
\begin{eqnarray}
R_{\textrm{step@}\omega_{4s,3s}/2} & = & \frac{P_{4s,\textrm{step@}\omega_{4s,3s}/2}}{P_{4s,\textrm{TL}}} \label{eq:R} \\
& = & \frac{\left|1 + \Im\left[A_{4s,\textrm{step@}\omega_{4s,3s}/2}^{(4)}\right]/A_{4s,*}^{(2)}\right|^{2}+
\left|\Re\left[A_{4s,\textrm{step@}\omega_{4s,3s}/2}^{(4)}\right]/A_{4s,*}^{(2)}\right|^{2}}
{\left|1 + \Im\left[A_{4s,\textrm{TL}}^{(4)}\right]/A_{4s,*}^{(2)}\right|^{2}+
\left|\Re\left[A_{4s,\textrm{TL}}^{(4)}\right]/A_{4s,*}^{(2)}\right|^{2}}
\nonumber \\
& = &
\frac{\left|1 + K_{\textrm{step@}\omega_{4s,3s}/2} I_{0} \right|^{2} + \left| \kappa I_{0} \right|^{2}}
     {\left|1 + K_{\textrm{TL}} I_{0} \right|^{2} + \left| \kappa I_{0} \right|^{2}} \nonumber \; ,
\end{eqnarray}
where the pulse shape dependence enters only via the $K$ factors.
Accounting for the magnitude of $\left|A_{4s,\textrm{TL}}^{(4)\textrm{near-res}}\right|$
vs.~$\left|A_{4s,\textrm{step@}\omega_{4s,3s}/2}^{(4)\textrm{near-res}}\right|$
and for the sign of $A_{4s}^{(4)\textrm{near-res}}$ relative to $A_{4s}^{(2)}$,   
one obtains the following behavior for the different $\lambda_{0}$ cases:
(i) For blue detuning of $\omega_{0} > \omega_{4s,3s}/2$ ($\lambda_{0}$=773.5~nm):
$K_{\textrm{step@}\omega_{4s,3s}/2} > 0$, $K_{\textrm{TL}} > 0$,
$\left|K_{\textrm{step@}\omega_{4s,3s}/2}\right| < \left|K_{\textrm{TL}}\right|$,
and thus $R_{\textrm{step@}\omega_{4s,3s}/2} < 1$;
(ii) For no detuning of $\omega_{0} = \omega_{4s,3s}/2$ ($\lambda_{0}$=777~nm):
$K_{\textrm{step@}\omega_{4s,3s}/2} = 0$, $K_{\textrm{TL}} = 0$,
and thus $R_{\textrm{step@}\omega_{4s,3s}/2} = 1$;
(iii) For red detuning of $\omega_{0} < \omega_{4s,3s}/2$ ($\lambda_{0}$=780~nm):
$K_{\textrm{step@}\omega_{4s,3s}/2} < 0$, $K_{\textrm{TL}} < 0$,
$\left|K_{\textrm{step@}\omega_{4s,3s}/2}\right| < \left|K_{\textrm{TL}}\right|$,
and thus $R_{\textrm{step@}\omega_{4s,3s}/2} > 1$.
In other words, the two-photon absorption induced by the shaped pulse of $\omega_{step} = \omega_{4s,3s}/2$
is lower, equal, or higher than the TL absorption 
according to whether $\omega_{0}$ ($\lambda_{0}$) is
blue-detuned, non-detuned, or red-detuned from $\omega_{4s,3s}/2$, respectively.
One can also see that, when $\left|R_{\textrm{step@}\omega_{4s,3s}/2}\right| \ne 1$, the value of
$\left|R_{\textrm{step@}\omega_{4s,3s}/2}\right|$ increases as $I_{0}$ increases.
This entire intermediate-field behavior is indeed the one observed in the results of Fig.~\ref{fig_3},
except for a small deviation of $R_{\textrm{step@}\omega_{4s,3s}/2}$ from a value of one in the case
of $\lambda_{0}$=777~nm [for example, in Fig.~\ref{fig_3}(2c) it reaches a value of 1.15].
This deviation originates from the resonance-mediated Raman term $A^{(\textrm{res}R)}$
that is excluded from $A^{(R)}$ in this part of the analysis
and is not symmetric around $\delta =0$ [as $A^{(\textrm{non-res}R)}(\delta)$ is].

\subsection{Intermediate-field two-photon absorption: shaped pulse with $\omega_{step} \approx \omega_{0}$}
\label{subsec3:discussion}

The third-row panels [panels (1c) and (3c)] of Fig.~\ref{fig_5} correspond to the shaped pulse
with $\omega_{step} = \omega_{0}$ in the detuned $\lambda_{0}$ cases of 773.5~nm and 780~nm
(the corresponding case with $\lambda_{0}$=777~nm is actually the one already considered
in Sec.~\ref{subsec2:discussion}).
It is considered here as the representative case for the region of $\omega_{step} \approx \omega_{0}$,
where a systematic attenuation or enhancement of the intermediate-field TL-normalized absorption relative to
the weak-field TL-normalized absorption 
occur [Figs.~\ref{fig_3}(1c) and (3c), Figs.~\ref{fig_4}(a) and (c)].
The attenuation or enhancement correspond, respectively, to the blue ($\lambda_{0}$=773.5~nm)
or red detuning ($\lambda_{0}$=780~nm) of $\omega_{0}$ from $\omega_{4s,3s}/2$.
Detailed analysis of the present case, as conducted above for the TL
pulse and for the shaped pulse with $\omega_{step} =
\omega_{4s,3s}/2$, leads to the conclusion that the only difference
here is that $A_{4s,\textrm{step@}\omega_{0}}^{(2)} \ne
A_{4s,\textrm{TL}}^{(2)}$ while
$A_{4s,\textrm{step@}\omega_{4s,3s}/2}^{(2)} =
A_{4s,\textrm{TL}}^{(2)}$.
All the other qualitative conclusions regarding the different amplitude components in terms of their magnitude,
sign, and real/imaginary character are exactly the same for both the
$\omega_{step} = \omega_{0}$ and $\omega_{step} = \omega_{4s,3s}/2$ pulses.

In general,
the ratio $RR_{\textrm{step}}$ between the intermediate-field and weak-field TL-normalized
absorption corresponding to the shaped pulse of a given $\omega_{step}$ is given by
\begin{eqnarray}
RR_{\textrm{step@}\omega_{step}} & = &
\frac{R^{\textrm{interm-field}}_{\textrm{step@}\omega_{step}}}{R^{\textrm{weak-field}}_{\textrm{step@}\omega_{step}}} \label{eq:RR} \\
& = &
\frac{\left|1 + \Im\left[A_{4s,\textrm{step@}\omega_{step}}^{(4)}\right]/A_{4s,\textrm{step@}\omega_{step}}^{(2)}\right|^{2}+
\left|\Re\left[A_{4s,\textrm{step@}\omega_{step}}^{(4)}\right]/A_{4s,\textrm{step@}\omega_{step}}^{(2)}\right|^{2}}
{\left|1 + \Im\left[A_{4s,\textrm{TL}}^{(4)}\right]/A_{4s,\textrm{TL}}^{(2)}\right|^{2}+
\left|\Re\left[A_{4s,\textrm{TL}}^{(4)}\right]/A_{4s,\textrm{TL}}^{(2)}\right|^{2}}
\nonumber \\
& = &
\frac{\left|1 + K_{\textrm{step@}\omega_{step}} I_{0} \right|^{2} + \left| \kappa I_{0} \right|^{2}}
     {\left|1 + K_{\textrm{TL}} I_{0} \right|^{2} + \left| \kappa I_{0} \right|^{2}}
\nonumber \; ,
\end{eqnarray}
where the pulse shape dependence enters only via the $K$ factors.
As can be seen, there is a close similarity between this equation
and Eq.~(\ref{eq:R}). So, based on the above analysis conclusions,
one gets the observed intermediate-field behavior for $\omega_{step}
= \omega_{0}$:
(i) For blue detuning of $\omega_{0} > \omega_{4s,3s}/2$ ($\lambda_{0}$=773.5~nm):
$K_{\textrm{step@}\omega_{0}} > 0$, $K_{\textrm{TL}} > 0$,
$\left|K_{\textrm{step@}\omega_{0}}\right| < \left|K_{\textrm{TL}}\right|$,
and thus $RR_{\textrm{step@}\omega_{0}} < 1$ (attenuation);
(ii) For red detuning of $\omega_{0} < \omega_{4s,3s}/2$ ($\lambda_{0}$=780~nm):
$K_{\textrm{step@}\omega_{0}} < 0$, $K_{\textrm{TL}} < 0$,
$\left|K_{\textrm{step@}\omega_{0}}\right| < \left|K_{\textrm{TL}}\right|$,
and thus $RR_{\textrm{step@}\omega_{0}} > 1$ (enhancement).
Also, for both $\lambda_{0}$ values,
the value of $\left|RR_{\textrm{step@}\omega_{0}}\right|$ increases as $I_{0}$ increases.
In the red detuning case ($\lambda_{0}$=780~nm), this increase leads eventually to an intermediate-field absorption
that exceeds the intermediate-field TL absorption [Fig.~\ref{fig_3}(3c)].
It worth emphasizing that the intermediate-field attenuation/enahncement effect considered here is relative
to the weak-field case of the same shaped pulse, while for the pulse with $\omega_{step} = \omega_{4s,3s}/2$
it is considered above relative to the intermediate-field TL absorption.


\subsection{Intermediate-field two-photon absorption: shaped pulse with $\omega_{step}$=$\omega_{7p,4s}$}

\begin{figure} 
\includegraphics[width=10cm]{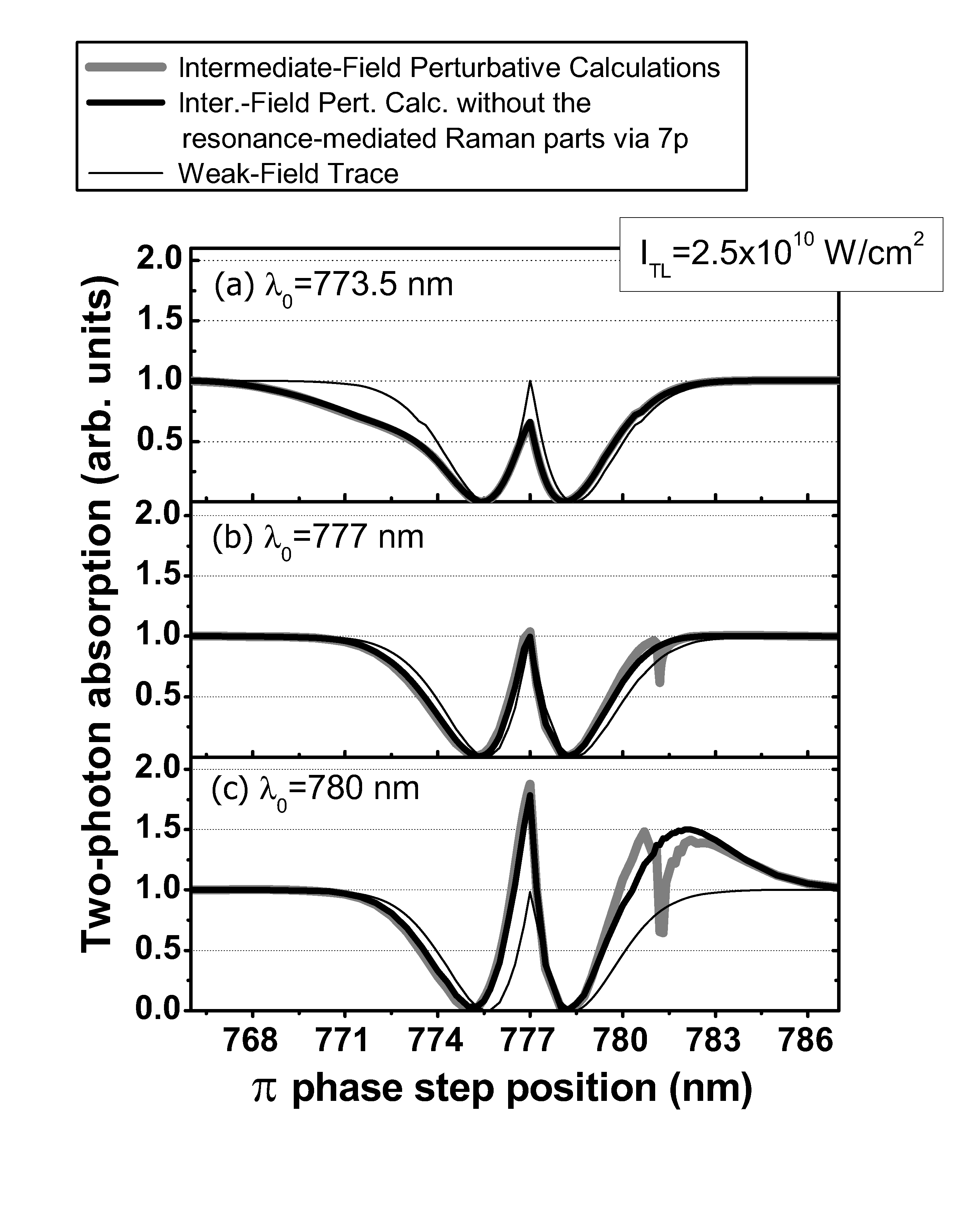}     
\caption{
Perturbative theoretical results (of 4$^{th}$ order)
for the two-photon absorption in Na induced by the shaped pulses having a $\pi$-step spectral phase pattern.
The traces show the final $4s$ population $P_{4s}$ as a function of the step position $\omega_{step}$.
The value of $P_{4s}$ is normalized by the final population $P_{4s,\textrm{TL}}$ excited by the corresponding transform-limited (TL) pulse.
Each panel corresponds to a different central spectral wavelength $\lambda_{0}$=773.5, 777, and 780~nm.
The traces given in gray thick lines are the real traces (shown also in Fig.~\ref{fig_3}) at
$I_{\scriptsize{\textrm{TL}}}= 2.5 \times 10^{10}$~W/cm$^{2}$, which is the upper
intensity limit of the present intermediate-field regime.
The traces given in black thick lines (without the intermediate-field dip feature at 781.2~nm)
are traces that have been calculated at $I_{\scriptsize{\textrm{TL}}}= 2.5 \times 10^{10}$~W/cm$^{2}$
with artificially setting to zero the contribution from the resonance-mediated Raman transitions via $7p$,
i.e., $A^{(\textrm{res}R)}=0$ (see text).
For completeness, also presented are the real weak-field traces
(black thin lines; also shown in Fig.~\ref{fig_3}). } \label{fig_4}
\end{figure}

As discussed above with regard to Fig.~\ref{fig_4},
the intermediate-field dip feature at $\omega_{step} = \omega_{7p,4s}$ (781.2~nm) originates from
the inclusion of the Raman term $A^{(\textrm{res}R)}$ in the
complete Raman term $A^{(R)}$ [Eq.~(\ref{eq8:amp-4th-order})-(\ref{eq10:amp-4th-order})].
As described above, $A^{(\textrm{res}R)}$ interferes only those Raman parts that are resonance-mediated via $7p$.
Each of them is either on resonance or near resonance with $7p$ ($\delta'$ is the corresponding detuning).
So, overall, each corresponding four-photon pathway is either on- or near-resonance with $4s$ and
either on- or near-resonance with $7p$.
The higher is the field amplitude $\left|\widetilde{E}(\omega_{7p,4s})\right|$ at $\omega_{7p,4s}$
the more prominent is the dip. So, here, it is most prominent for $\lambda_{0}$=780~nm [Fig.~\ref{fig_4}(c)].

With the inclusion of $A^{(\textrm{res}R)}$, the terms
$A_{4s}^{(4)\textrm{near-res}}$ and $A_{4s}^{(4)\textrm{on-res}}$ become complex quantities,
and thus they both contribute to both the imaginary and real parts of $A_{4s}^{(4)}$.
So, the analysis of the corresponding interference mechanism is much more complicated
as compared to the above, when $A^{(\textrm{res}R)}$ is excluded.
Essentially, the interference mechanism leading to the dip is very similar to the one we recently
identified in weak-field resonance-mediated (2+1) three-photon absorption \cite{amitay_3ph_2plus1}, where
a $\pi$-step at $\omega_{7p,4s}$ leads to a strong enhancement in the population transfer to the $7p$ state.
Here, for example, a $\pi$-step at $\omega_{7p,4s}$ leads to a constructive add-up within
the $\wp$-integral of $A_{4s}^{(4)\textrm{near-res}}$ [Eq.~(\ref{eq7.3:amp-4th-order})] between
the amplitudes contributed by the four-photon pathways on-resonant with $7p$ that are
of positive detuning $+$$\left|\delta\right|$ and of negative detuning $-$$\left|\delta\right|$ from $4s$.
The corresponding Raman transition involves the absorption of the photon $\omega_{7p,4s}\pm\delta$
and the emission of the photon $\omega_{7p,4s}$.
Since $\widetilde{E}(\omega_{7p,4s}+|\delta|)$ and
$\widetilde{E}(\omega_{7p,4s}-|\delta|)$ are of opposite signs for
the shaped pulse of $\omega_{step} = \omega_{7p,4s}$, the sign of
the detuning $\delta$ (and of the integrand factor $1/\delta$)
becomes correlated with the sign of $\widetilde{E}(\omega_{7p,4s}+\delta)$,
leading to the constructive add-up and resulting amplitude enhancement.
For the TL pulse, the add-up is destructive since the signs of
$\widetilde{E}(\omega_{7p,4s}-|\delta|)$ and $\widetilde{E}(\omega_{7p,4s}+|\delta|)$
are the same.
Upon a detailed analysis,
accounting also for the $\delta$-dependence of $A^{(2)}(\omega_{4s,3s}-\left|\delta\right|)$
when $\omega_{step}=\omega_{7p,4s}$ (781.2~nm)
(very similar to the one when $\lambda_{step}$=780~nm [Fig.~\ref{fig_5}(3c)]),
one obtains that, for the present Na excitation,
the sign of the amplitude contributed to $A_{4s}^{(4)}$ by the four-photon pathways
with non-resonant Raman parts [i.e., corresponding to $A^{(\textrm{res}R)}$]
is opposite to the sign of the amplitude contributed 
by those with resonance-mediated Raman parts [i.e., corresponding to $A^{(\textrm{res}R)}$].
This leads to a dip feature rather than a peak feature.

\section{\label{sec:Conclusions}Conclusions}
In conclusion, coherent control of femtosecond two-photon absorption in the intermediate-field regime
in analyzed in detail using a powerful frequency domain description that is based
on 4$^{th}$-order perturbation theory.
The two-photon absorption is coherently induced by non-resonant two-photon transitions as well as
by four-photon transitions that introduce a resonance-mediated nature to the excitation.
Their relative contributions to the total absorption amplitude depend on the field strength.
The corresponding interference mechanism is identified to include intra-group and inter-group interferences
involving these two groups of multiphoton transitions.
The inter-group interferences lead a difference between the
intermediate-field and weak-field absorption dynamics.
Their constructive/destructive nature 
is found to depend on the detuning direction of the pulse spectrum
from half the two-photon transition
frequency; It changes upon a red-to-blue detuning change. 
The extended frequency-domain description and its detailed
understanding serve as a basis for femtosecond control with
rationally-shaped pulses in a regime of significant absorption
yields, reaching population transfer in the range of 10-40\%
(depending on the specific system and excitation scheme).
They also serve as a basis for future extensions to molecular systems, to
other types of multiphoton processes, and to more complicated excitation schemes.


\section*{ACKNOWLEDGMENTS}
This research was supported by The Israel Science Foundation (grant No. 127/02),
by The James Franck Program in Laser Matter Interaction,
and by The Technion's Fund for The Promotion of Research.





\newpage

\newpage

\newpage

\newpage

\newpage

\end{document}